\documentclass{ws-mpla}
\usepackage[super,compress]{cite}
\usepackage{graphicx}
\begin{document}
\markboth{Hideki Asada}{Gravitational lensing by exotic objects}

%
\catchline{}{}{}{}{}
%

\title{Gravitational lensing by exotic objects}

\author{Hideki Asada}

\address{Faculty of Science and Technology, Hirosaki University,
Aomori 036-8561, Japan\\
asada@hirosaki-u.ac.jp}

%

\maketitle

\begin{history}
\received{Day Month Year}
\revised{Day Month Year}
\end{history}

\begin{abstract}
This paper reviews a phenomenological 
approach to the gravitational lensing by exotic objects 
such as the Ellis wormhole lens, 
where exotic lens objects may follow 
a non-standard form of the equation of state 
or may obey a modified gravity theory. 
A gravitational lens model is proposed 
in the inverse powers of the distance, 
such that the Schwarzschild lens 
and exotic lenses
can be described in a unified manner 
as a one parameter family. 
As observational implications, 
the magnification, shear, photo-centroid motion 
and time delay in this lens model are discussed.   
\end{abstract}

\ccode{PACS numbers: 04.40.-b, 95.30.Sf, 98.62.Sb}


\section{Introduction}	
General relativistic effects on the light propagation provide 
powerful tools in modern astronomy and cosmology. 
The gravitational lens effects enable us to study 
not only stars and galaxies but also 
extrasolar planets, dark matter and dark energy. 
Very recently, the first detection of the astrometric microlensing 
was successfully made to measure the mass of the white dwarf Stein 2051 B 
by using the Hubble space telescope \cite{Sahu}. 

Can we use the gravitational lens effects 
for the purpose of probing other exotic objects? 
Exotic objects may give us an opportunity 
for examining new physics such as 
a non-standard equation of state of the matter 
and a modified gravity theory. 

Also on the theoretical side, 
it is of great interest to probe such a new object
by using the gravitational lens effects. 
For instance, a rigorous way of constructing the lens equation 
with defining the bending angle of light,  
especially in a strong gravitational field, is argued 
\cite{Ohanian, Frittelli, VE2000, Virbhadra, VNC, VE2002, VK2008, 
ERT, Perlick}. 
Please see also Ref. \citen{Bozza2008} 
for a comparison of these different lens equations. 
Strong gravitational lensing was studied 
for a Schwarzschild black hole \cite{Frittelli, VE2000, Virbhadra}, 
for some naked singularities \cite{VE2002, VK2008}, 
for Reissner-Nordstr\"om black holes \cite{ERT},  
for Barriola-Vilenkin monopole \cite{Perlick} 
and 
for an Ellis wormhole \cite{CC, Clement,Ellis, Morris1, Morris2}. 
In particular, Ellis wormhole lenses have attracted renewed interests 
as an observational probe of violations of some energy conditions 
\cite{Perlick, Visser, Safonova, Shatskii, Nandi, Abe, Toki, Tsukamoto, Tsukamoto2013, Yoo, DS, BP, Nakajima, Gibbons}. 

Even a small deviation from the theory of general relativity can make 
a significant change in the gravitational lensing. 
For instance, a fourth order $f(R)$ gravity theory 
and the associated gravitational lensing 
were studied by several authors 
(e.g. \cite{Capozziello, Horvath, Mendoza}). 
However, 
the total magnification of the lensed images 
is stable and always larger than the unity 
against a small spherical perturbation of 
the Schwarzschild lens \cite{Asada2011}. 
Therefore, a perturbative approach around the Schwarzschild lens 
may be useful even for studying the gravitational lens 
in such a modified gravity theory. 

In addition, the gravitational lensing in higher dimensions 
is interesting. 
For instance, the gravitational lensing in the Tangherlini spacetime, 
which is an extension of Schwarzschild solution  
to arbitrary dimensions, was studied 
\cite{Tsukamoto2014}. 
There are the so-called tidal charges 
in the Dadhich et al. solution \cite{Dadhich}. 
The lens effects due to negative convergence 
by the tidal charges were investigated \cite{HGH}, 
where the tidal charge of a brane world black hole 
is caused by tidal forces acting on the brane-bulk boundary. 
Interestingly, the negative convergence in this case 
does not need any exotic matter. 
It originates from the Weyl curvature in higher dimensions. 
Note that asymptotically flat curvature-regular solitons 
with negative mass 
can belong to a class of the Myers-Perry family, 
though the soliton solutions in the odd spacetime dimensions 
are unlikely to have a relation with astrophysical objects \cite{GK}. 
In addition, traversable wormholes have been thought to need exotic 
matter or energy. 
According to a recent study, however, 
they can be constructed in the dilatonic Einstein-Gauss-Bonnet theory 
without using any form of exotic matter 
\cite{Kunz2011, Kunz2012}. 
These recent works provide us an increasing motivation 
for studying the gravitational lensing by exotic matter (or energy) 
or by new physics beyond the standard theory. 

The purpose of this paper is to review a phenomenological approach 
to gravitational lensing by exotic objects in the weak field approximation 
\cite{Kitamura2013,Izumi,Kitamura2014,Nakajima2014}. 
The present paper mentions 
some observational implications by this exotic lens model (KNA model). 
The proposed spacetime metric depends on the inverse distance 
to the power of positive $n$. 
This inverse power metric was proposed 
by Kitamura and his collaborators \cite{Kitamura2013}. 
They started from the phenomenological spacetime metric 
to obtain the weak deflection angle of light 
in the inverse power of the impact parameter. 
This deflection angle of light 
is a simple extension of that by the Schwarzschild lens. 
The same form of the weak deflection angle of light 
was assumed as an ansatz independently by another group
\cite{Tsukamoto2013}.  
They used the modified deflection angle of light 
to study whether the Ellis wormhole lens can be 
distinguished from the Schwarzschild lens. 
A similar inverse-power law was used for modeling 
not only a single lens but also a binary lens 
\cite{Bozza2015,Bozza2016,Bozza2017}. 

Throughout this paper, we assume the asymptotic source and observer.  
See e.g. Refs. \citen{Ishihara2016, Ishihara2017} 
for computations of the bending angle of light 
for the finite distance between the source and the observer. 
We assume also the static observer.

\section{Ellis Wormhole}
\subsection{Gravitational deflection angle of light by Ellis wormholes}
We begin with the Ellis wormhole lens 
as an example of gravitational deflection of light by exotic objects, 
The spacetime metric for the Ellis wormhole can be written as 
\cite{Ellis, Perlick, Nandi}
\begin{equation}
ds^2 = - dt^2 + dr^2 + (r^2 + a^2) (d\theta^2 + \sin^2\theta d\phi^2) . 
\label{ds-Ellis}
\end{equation}
To fully cover the wormhole geometry, the radial coordinate $r$ 
takes from $-\infty$ to $+\infty$, 
where $r=0$ means the throat of the wormhole. 
This throat often make confusions in the literature 
as explained later. 
We shall discuss the deflection angle of light. 
For this purpose, it is sufficient for us to consider 
only one half of the wormhole geometry. 
Without loss of generality, we choose 
$r \in (0, +\infty)$. 
We consider 
the Lagrangian for a massless (light-like) particle 
traveling in the Ellis wormhole spacetime. 
It becomes 
\begin{equation}
L = - \dot{t}^2 + \dot{r}^2 
+ (r^2 + a^2) (\dot{\theta}^2 + \sin^2\theta \dot{\phi}^2) , 
\label{L}
\end{equation}
where the dot is the derivative with respect to 
the affine parameter along the light ray. 

The Ellis wormhole is spherically symmetric. 
Without loss of generality, therefore, 
we can choose the orbital plane 
as the equatorial plane $\theta = \pi/2$. 

Moreover, 
the Ellis wormhole spacetime has the time-like Killing vector 
and the rotational Killing vector. 
Therefore, there are two constants of motion of a photon. 
The two constants can be defined as 
\begin{eqnarray}
E &\equiv& \dot{t} , 
\label{E}
\\
h &\equiv& (r^2+a^2)\dot{\phi} . 
\label{h}
\end{eqnarray} 
Here, $E$ and $h$ denote the photon's specific energy 
and specific angular momentum, respectively. 
We substitute the two constants of motion into the null condition 
$ds^2 = 0$ 
to find an equation for the photon orbit. 
The photon orbit equation is obtained as 
\begin{equation}
\frac{1}{(r^2+a^2)^2} \left( \frac{dr}{d\phi} \right)^2 
= \frac{1}{b^2} - \frac{1}{r^2+a^2} . 
\label{Ellis-orbit}
\end{equation}
Here, $b \equiv h/E$ denotes the impact parameter. 

In general, the impact parameter is the perpendicular coordinate distance 
between the fiducial path and the center of a deflector, 
where we assume that the fiducial path were not affected by the deflector. 
For the case of the Ellis wormhole, this zero deflection limit 
is achieved by the limit as $a \to 0$. 
If $a = 0$, $r=b$ means that $r$ is the minimum 
according to Eq. (\ref{Ellis-orbit}). 
Hence, the above constant $b$ can be safely called the impact parameter 
of the light trajectory. 

On the other hand, the closest approach $r_0$ 
is the distance 
between the photon orbit (not the fiducial path) 
and the center of the deflector. 
The closest approach is given by Eq. (\ref{Ellis-orbit}) as 
\begin{equation}
r_0 = \sqrt{b^2 - a^2} . 
\label{r0}
\end{equation} 
Namely, $r_0$ is the minimum value of the radial coordinate 
on the light ray. 

We integrate Eq. (\ref{Ellis-orbit}) to obtain 
the deflection angle as 
\begin{equation}
\alpha(b) = 2 \int_{r_0}^{\infty} 
\frac{bdr}{\sqrt{(r^2+a^2)^2 - (r^2+a^2)b^2}} 
- \pi . 
\label{alpha-r}
\end{equation}
For the later convenience, 
we make a coordinate transformation from $r \in [0, +\infty)$ 
to $R \in [a, +\infty)$ by defining $R^2 = r^2+a^2$, 
where 
$R$ means the circumference radius. 
Eq. (\ref{alpha-r}) becomes  
\begin{equation}
\alpha(b) = 2 \int_{b}^{\infty} 
\frac{bdR}{\sqrt{(R^2-a^2) (R^2-b^2)}} 
- \pi . 
\label{alpha-R}
\end{equation}
This can be rewritten as 
\begin{eqnarray}
\alpha(b) &=& 2 \int_0^1 \frac{dt}{\sqrt{(1-t^2)(1 - k^2 t^2)}} 
- \pi 
\nonumber\\
&=& 2 K(k) - \pi , 
\label{alpha2}
\end{eqnarray}
where we define $t \equiv b/R$, $k \equiv a/b$ 
and we denote $K(k)$ as the complete elliptic integral of the first kind. 
This integral form allows us to immediately obtain 
a Taylor series expansion for $k <1$ as 
\begin{equation}
\alpha(b) = \pi \sum_{n=1}^{\infty} 
\left[ \frac{(2n-1)!!}{(2n)!!} \right]^2 k^{2n} . 
\label{alpha3}
\end{equation}

Perlick \cite{Perlick} and Nandi, Zhang and Zakharov \cite{Nandi}
obtained the deflection angle in a different form 
(e.g., Eq. (54) in \cite{Nandi}) 
that is expressed in terms of the closest approach 
\cite{Perlick, Nandi}. 
Clearly, their expression with using the closest approach 
can be recovered from Eq. (\ref{alpha2}) 
by noting the relation $r_0^2 = b^2 - a^2$ \cite{Nandi-2}. 
The expression by Eq. (\ref{alpha2}) seems 
more useful for astronomers, 
especially on a microlens study, 
because describing an image direction (its angular position) 
needs the impact parameter but not the closest approach. 

Dey and Sen \cite{DS} employed an alternative method 
that was proposed by Amore and Arceo \cite{Amore, Amore2}, 
In this method, firstly,  
the linear delta function technique is used, 
such that an integral with an {\it ansatz} potential 
can be approximated. 
Next, the principle of minimal sensitivity (PMS) is used to 
minimize the parametric dependence on the deflection angle. 
As a result, they obtained the deflection angle by the Ellis wormhole 
as 
\begin{equation}
\alpha = \pi \left\{\sqrt 
\frac{2 (r_0^2 + a^2)}{2 r_0^2 + a^2} -1 \right\} .
\label{alpha-DS}
\end{equation}
In the weak field approximation ($a \ll b \sim r_0$), 
their deflection angle is expressed in the Taylor-series as 
\begin{equation}
\alpha = \frac{\pi}{4} \left(\frac{a}{r_0}\right)^2 
- \frac{5 \pi}{32} \left(\frac{a}{r_0}\right)^4 
+ O\left(\frac{a}{r_0}\right)^6 . 
\label{alpha-DS2}
\end{equation}
In terms of the impact parameter, 
Eq. (\ref{alpha-DS2}) can be rearranged as 
\begin{equation}
\alpha(b) = \frac{\pi}{4} \left(\frac{a}{b}\right)^2
+ \frac{3 \pi}{32} \left(\frac{a}{b}\right)^4
+ O\left(\frac{a}{b}\right)^6 . 
\label{alpha-DS3}
\end{equation}
where we used $r_0^2 = b^2 - a^2$. 

Eq. (\ref{alpha2}), that is the rigorous form of the deflection angle,  
is expanded in the weak field as 
\begin{equation}
\alpha(b) = \frac{\pi}{4} \left(\frac{a}{b}\right)^2
+ \frac{9 \pi}{64} \left(\frac{a}{b}\right)^4 
+ O\left(\frac{a}{b}\right)^6 . 
\label{alpha-approx}
\end{equation}

Eq. (\ref{alpha-DS3}) recovers the first term in
Eq. (\ref{alpha-approx}), 
though it misses the second (and more) terms. 

Let us discuss why the PMS method fails for the Ellis wormhole.  
The Schwarzschild spacetime has a singularity at $r=0$, 
which leads to a divergent behavior of the light bending 
near the singularity. 
Therefore, the PMS approximation using the delta function works 
\cite{Amore, Amore2}. 
However, $r=0$ in the Ellis geometry is not a singularity. 
It is a {\it regular} sphere 
which can connect with a separate spatial region. 
Therefore, the deflection angle by the Ellis wormhole is not inversely 
but logarithmically divergent around $r=0$. 
This is why the PMS does not well work for the Ellis wormhole. 

As a practical example, 
we consider a particular case that the closest approach vanishes 
($r_0 = 0$), namely $b=a$. 
From Eq. (\ref{alpha2}), we arrive at 
\begin{eqnarray}
\alpha(a) &=& 2 \int_0^1 \frac{dt}{1-t^2}
- \pi 
\nonumber\\
&\sim& \ln\infty . 
\label{alpha-a}
\end{eqnarray}
The throat $r=0$ means a light sphere (photon sphere), 
on which a light ray can eternally stay if it is tangential 
to the sphere. 
This can be understood by seeing that 
$r=0$ satisfies Eq. (\ref{Ellis-orbit}). 
The logarithmic divergence in Eq. (\ref{alpha-a}) 
can be thus related with the existence of the light sphere. 
On the other hand, Eq. (\ref{alpha-DS}) becomes 
$\alpha \to \pi (\sqrt2 -1)$ as $r_0 \to 0$ ($b \to a$). 
This result is incorrectly finite, 
because the throat effects are not taken into account by this approach. 
 
Strong deflection limit in the Ellis wormhole and its corrections 
to microlens light curves were discussed by Tsukamoto 
\cite{Tsukamoto2016}. 
Light curves for light bundles traveling 
through the Ellis wormhole throat were investigated 
\cite{TH2017}. 
The so-called retro-lensing by the Ellis wormhole, 
which is a consequence of the reflection of light around the throat, 
was argued 
\cite{Tsukamoto2017}. 

\subsection{Observational constraints on wormholes}
Abe \cite{Abe} proposed a method for microlensing constraints on 
the wormholes that might exist quietly in our galaxy. 
He assumed the Ellis wormhole as a gravitational lens 
to study the wormhole microlensing. 
He concluded that, 
if the wormholes in our galaxy have a throat radius 
between $100$ and $10^7$ km and 
their number density is comparable to that of ordinary stars, 
their detection can be achieved in principle 
by reanalyzing past data sets in microlensing surveys. 
For such reanalysis, he discussed the microlensed light curve 
by the wormhole. 

The first detection of the astrometric microlens 
by a star, more precisely a white dwarf, was done by using 
the Hubble space telescope \cite{Sahu}.  
In the future, such a astrometric microlens technique 
will be used for probing exotic objects in our universe. 
Possible astrometric constraints on the Ellis wormhole 
were discussed by Toki et al. \cite{Toki}

Takahashi and Asada used an observational result 
in the Sloan Digital Sky Survey Quasar Lens Search (SQLS) \cite{TA}.  
They argued the first cosmological constraints 
on 
Ellis wormholes and negative-mass compact objects as follows. 
A key for the cosmological constraint is that 
there is no multiple image lensed by the above two 
exotic objects for 
$\sim 50000$ distant quasars in the SQLS data. 
Here, a large number of quasars mean a large survey volume. 
This null detection by SQLS suggests an upper bound 
on the abundances of these lenses over a cosmic survey volume. 
The number density of negative mass compact objects 
is constrained as 
$n<10^{-8} (10^{-4}) h^3 {\rm Mpc}^{-3}$ for the mass scale
 $|M| > 10^{15} (10^{12}) M_\odot$. 
This corresponds to a constraint on 
the cosmological density parameter 
$|\Omega| < 10^{-4}$ at the galaxy-scale mass range 
$|M|=10^{12-15}M_\odot$. 
The number density of the Ellis wormhole 
is constrained as 
$n<10^{-4} h^3 {\rm Mpc}^{-3}$ 
for the throat radius 
around $a = 10^{1-4}$pc. 
This range of the throat radius is much smaller than the Einstein ring radius. 
Hence, the weak field approximation holds in this analysis. 
Please see also 
Refs. \citen{Bondi, Jammer1961, Jammer1999, Cramer, Piran} 
for a review on fundamental aspects of negative mass.

\section{Introducing the Inverse-power model (KNA model)}
Kitamura, Nakajima and Asada \cite{Kitamura2013} 
proposed a very tractable form of the lens model by exotic objects. 
In particular, this model has an advantage that 
it can be easily treated even by hand. 
They suggested a usage of 
an asymptotically flat, static and spherically symmetric 
spacetime metric that depends on 
the inverse distance to the power of positive $n$ 
in the weak field approximation. 
They investigated the light propagation through a four-dimensional spacetime, 
though the whole spacetime may be higher dimensional. 
This is because the electromagnetic field is located only in 
the four-dimensional spacetime. 
The proposed four-dimensional spacetime metric (KNA model) is written as 
\begin{equation}
ds^2=-\left(1-\frac{\varepsilon_1}{r^n}\right)dt^2
+\left(1+\frac{\varepsilon_2}{r^n}\right)dr^2
+r^2(d\Theta^2+\sin^2\Theta d\phi^2) 
+O(\varepsilon_1^2, \varepsilon_2^2, \varepsilon_1 \varepsilon_2) ,  
\label{ds}
\end{equation}
where $r$ is the circumference radius and 
$\varepsilon_1$ and $\varepsilon_2$ are book-keeping 
parameters that are frequently used 
in the following iterative calculations. 
Here, $\varepsilon_1$ and $\varepsilon_2$ 
may be either positive or negative, respectively. 
For instance, 
the case that both $\varepsilon_1$ and $\varepsilon_2$ for $n=1$ 
are negative 
correspond to the linearized Schwarzschild metric 
with negative mass. 
Note that the above proposed metric is used 
only for the weak gravitational region. 
Its extension to the strong field such as 
the neighborhood of a black hole horizon is left as a future work.

\section{Deflection angle and convergence}
As usual, 
some calculations on the light propagation 
can be considerably simplified 
by using a conformal transformation. 
This is because the conformal transformation does not influence 
the null structure such as the light propagation. 
In the present case, a conformal factor can be chosen 
as $(1-\varepsilon_1/r^n)^{1/2}$.

We define $\varepsilon \equiv n \varepsilon_1 + \varepsilon_2$ 
and 
\begin{equation}
R^2 \equiv \dfrac{r^2}{\left(1-\dfrac{\varepsilon _1}{r^n}\right)} . 
\label{R}
\end{equation}
At the linear order in $\varepsilon_1$ and $\varepsilon_2$, 
the proposed spacetime metric takes a simpler form as 
\begin{equation}
d\bar{s}^2=-dt^2+\left(1+\dfrac{\varepsilon}{R^n}\right)dR^2
+R^2 (d\theta^2+\sin^2\theta d\phi^2) 
+O(\varepsilon^2) . 
\label{d2}
\end{equation}
There is the only one parameter $\varepsilon$ in 
the conformally transformed metric. 
Hence, some calculations can be much simpler. 

Without loss of generality, we focus on 
the equatorial plane $\Theta = \pi/2$, 
because of the spherical symmetry of the spacetime. 
The deflection angle of light at the linear order in $\varepsilon$ 
is computed as \cite{Kitamura2013} 
\begin{align}
\alpha
&=\dfrac{\varepsilon}{b^n}\int_0^{\frac{\pi}{2}} \cos^n\psi d\psi 
+O(\varepsilon^2) .  
\label{alpha}
\end{align}
Here, the integral is a positive constant, 
$b$ denotes the impact parameter of the light ray. 
We absorb the positive constant 
into the parameter $\varepsilon$, 
such that the linear-order deflection angle of light 
can be written simply as 
$\alpha = \bar\varepsilon/b^n$. 
Note that the sign of $\bar\varepsilon$ is the same as 
that of $\varepsilon$. 
This deflection angle at the linear order recovers both 
the Schwarzschild ($n=1$) and Ellis wormhole ($n=2$) cases. 
This means that the propose spacetime metric is 
a one-parameter family that can accommodate the Schwarzschild 
and the Ellis wormhole metrics in the weak field approximation. 

For $\varepsilon > 0$, the deflection angle of light 
in the proposed spacetime model is always positive. 
This means the gravitational pull on light rays 
in the corresponding spacetime model. 
For $\varepsilon < 0$, on the other hand, 
the deflection angle of light is always negative.  
This implies the gravitational repulsion on light rays, 
which has an analogy with a concave lens in optics. 
Tsukamoto and Harada \cite{Tsukamoto2013} employed, 
as an {\it ansatz}, 
the modified bending angle same as what is derived above 
from the spacetime metric. 
Kitamura et al. \cite{Kitamura2013} derived the modified bending angle
of light by starting from the inverse-power spacetime metric. 

Before closing this section, 
we discuss a connection with the present lens model 
and an effective mass. 
We consider a simple application of the standard lens theory \cite{SEF}, 
though the validity of the application is not clear 
for the exotic lens model so far. 
By this application, we can see that the deflection angle of light 
in the form of 
$\alpha = \bar\varepsilon/b^n$ 
shows an interesting relation 
with a convergence (corresponding to scaled surface mass density) as 
\begin{equation}
\kappa(b) = \frac{\bar\varepsilon (1-n)}{2} \frac{1}{b^{n+1}} . 
\label{kappa}
\end{equation}

For the linearized Schwarzschild case ($n = 1$), for instance, 
the convergence vanishes everywhere. 
In the framework of the standard lens theory, 
the effective surface mass density of the lens object 
with $\varepsilon > 0$ and $n>1$ 
can be interpreted as {\it negative}  \cite{Kitamura2013}. 
Namely, the matter (and energy) need to be exotic 
for this case. 
Also for $\varepsilon < 0$ and $n<1$, 
the convergence is negative and hence 
the matter (and energy) need to be exotic. 
More interestingly, the convergence for $\varepsilon < 0$ and $n>1$ 
is positive everywhere except for the central singularity of the
spacetime, though the weak field approximation breaks down 
in the vicinity of the singularity. 
Hence, exotic matter (and energy) are not required 
according to the framework of the standard lens theory, 
in spite of the gravitational repulsion on light rays. 
Note that 
attraction ($\varepsilon > 0$) and repulsion ($\varepsilon < 0$) 
in the above discussions are not always one-to-one correspondence 
to positive convergence $\kappa > 0$ and negative one $\kappa < 0$.

\section{Lens equation and image positions}
In this section, let us discuss the lensing shear by exotic objects. 
Under the thin lens approximation, 
the lens equation is written as \cite{SEF} 
\begin{equation}
\beta = \frac{b}{D_{\rm{L}}} - \frac{D_{\rm{LS}}}{D_{\rm{S}}} \alpha(b) . 
\label{lenseq}
\end{equation}
Here, we denote 
$\beta$, $D_L$, $D_S$ and $D_{LS}$ 
as the angular position of the source, 
the angular distance from the observer to the lens, 
that from the observer to the source, and 
that from the lens to the source, respectively. 
The angular position of the image is denoted as $\theta = b/D_L$. 

\begin{figure}
\includegraphics[width=6cm]{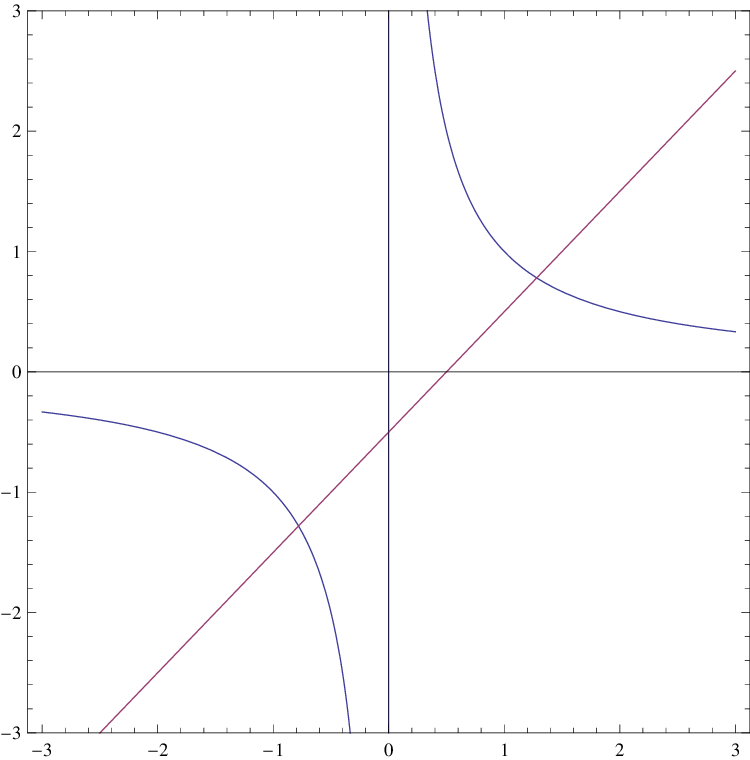}
\includegraphics[width=6cm]{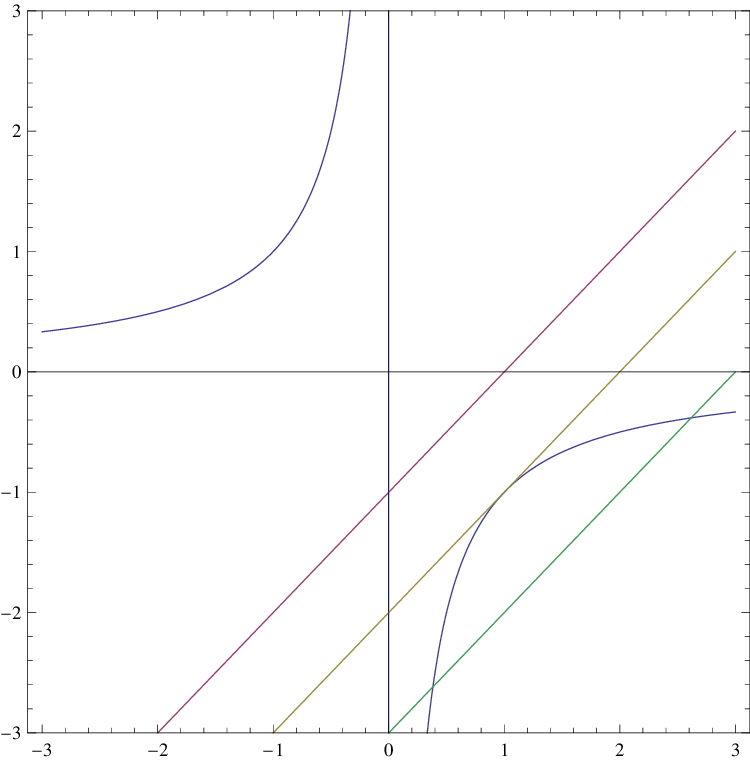}
\caption{
Schematic figure for the number of the lensed images 
(in arbitrary units). 
The left and right panels correspond to 
$\varepsilon >0$ and $\varepsilon < 0$, respectively. 
The curves denote $\pm 1/\theta^n$, 
where $+$ is for $\varepsilon >0$ and $-$ is for $\varepsilon < 0$. 
The straight lines denote $\theta - \beta$. 
The intersection points are positions of the lensed images. 
There are three cases for $\varepsilon < 0$: 
Zero image, the only one image and two images. 
}
\label{fig-image}
\end{figure}

For $\varepsilon > 0$, 
there always exists a positive root. 
This root is corresponding to 
the Einstein ring that is caused for the $\beta=0$ case. 
To be more precise, 
the Einstein ring radius is defined as \cite{SEF} 
\begin{equation}
\theta_{\rm{E}} \equiv 
\left(
\frac{\bar\varepsilon D_{\rm{LS}}}{D_{\rm{S}} D_{\rm{L}}^n}
\right)^{\frac{1}{n+1}} .
\label{theta_E}
\end{equation} 
If $\varepsilon < 0$, on the other hand, 
Eq. (\ref{lenseq}) has no positive root for $\beta = 0$. 
This is because 
this case describes the repulsive force 
and the light emitted by the source behind the lens object 
cannot reach the observer. 
See also Figure \ref{fig-image} for the number of the lensed images. 
For the later convenience, 
we define the (virtual) Einstein ring radius for $\varepsilon < 0$ 
as 
\begin{equation}
\theta_{\rm{E}} \equiv 
\left(
\frac{|\bar\varepsilon| D_{\rm{LS}}}{D_{\rm{S}} D_{\rm{L}}^n}
\right)^{\frac{1}{n+1}} , 
\label{theta_E2}
\end{equation} 
such that the lens equation can be rewritten 
in terms of the normalized variables. 
This virtual Einstein radius allows us to estimate a typical angular size 
for $\varepsilon < 0$ lenses, 
though the Einstein ring does not really exist for this case. 

\section{Shear}
In this section, 
we follow Ref. \citen{Izumi} to consider the gravitational lens shear 
by an exotic object. 

\subsection{$\varepsilon > 0$ case}
First, let us begin with a $\varepsilon > 0$ case. 
As mentioned above, the matter (and energy) need to be exotic if $n > 1$. 
In the units of the Einstein ring radius, 
Eq. (\ref{lenseq}) takes the vectorial form as 
\begin{eqnarray}
\boldsymbol{\hat\beta}
&=& \boldsymbol{\hat\theta} 
- \frac{\boldsymbol{\hat\theta}}{\hat\theta^{n+1}}  
\quad (\hat\theta > 0) , 
\label{lenseqP}\\
\boldsymbol{\hat\beta}
&=& \boldsymbol{\hat\theta} 
- \frac{\boldsymbol{\hat\theta}}{(-\hat\theta)^{n+1}}  
\quad (\hat\theta < 0) , 
\label{lenseqM}
\end{eqnarray}
where we define $\hat\beta \equiv \beta/\theta_E$ and 
$\hat\theta \equiv \theta/\theta_E$, 
and $\boldsymbol{\hat\beta}$ and $\boldsymbol{\hat\theta}$ 
denote the corresponding vectors. 
There is always one image in a region of $\hat\theta > 0$, 
whereas the other image appears in a region of $\hat\theta < 0$ 
\cite{Kitamura2013}. 

The magnification matrix can be defined as 
$A_{ij} \equiv \partial\beta^i/\partial\theta_j$ \cite{SEF}. 
By doing straightforward calculations, 
the magnification matrix for $\hat\theta > 0$ 
can be rewritten as  
\begin{eqnarray}
(A_{ij}) &=& 
\left(
\begin{array}{cc}
1 - \cfrac{1}{\hat\theta^{n+1}} 
+ (n+1) \cfrac{\hat\theta_x \hat\theta_x}{\hat\theta^{n+3}}
& (n+1) \cfrac{\hat\theta_x \hat\theta_y}{\hat\theta^{n+3}} \\
(n+1) \cfrac{\hat\theta_x \hat\theta_y}{\hat\theta^{n+3}} & 
1 - \cfrac{1}{\hat\theta^{n+1}} 
+ (n+1) \cfrac{\hat\theta_y \hat\theta_y}{\hat\theta^{n+3}} 
\end{array}
\right) . 
\label{Aij}
\end{eqnarray}
By using its eigen values $\lambda_{\pm}$, 
this matrix can be diagonalized as 
\begin{eqnarray}
(A_{ij}) &\equiv& 
\left(
\begin{array}{cc}
1-\kappa-\gamma & 0 \\
0 &  1-\kappa+\gamma 
\end{array}
\right) 
\nonumber\\
&\equiv& 
\left(
\begin{array}{cc}
\lambda_{-} & 0 \\
0 & \lambda_{+} 
\end{array}
\right) . 
\end{eqnarray}
Here, the $x$ and $y$ coordinates are chosen 
along the radial and tangential directions, respectively, 
such that $(\hat\theta_i) = (\hat\theta, 0)$ 
and $(\hat\beta_i) = (\hat\beta, 0)$. 
As a result, the radial magnification factor is $1/\lambda_{-}$, 
and the tangential one is $1/\lambda_{+}$. 

We study the primary image ($\hat\theta > 0$). 
By using Eq. (\ref{lenseqP}), we get 
\begin{eqnarray}
\lambda_{+} = \frac{\hat\beta}{\hat\theta} 
= 1 - \frac{1}{\hat\theta^{n+1}} , 
\label{lambdaP}
\end{eqnarray}
\begin{eqnarray}
\lambda_{-} = \frac{d\hat\beta}{d\hat\theta} 
= 1 + \frac{n}{\hat\theta^{n+1}} . 
\label{lambdaM}
\end{eqnarray}
 
We briefly mention a short cut 
for getting Eqs. (\ref{lambdaP}) and (\ref{lambdaM}). 
One may choose the $x$ and $y$ coordinates 
to be locally along the radial and tangential directions, 
respectively. 
Namely, $(\hat\theta_i) = (\hat\theta, 0)$ 
and $(\hat\beta_i) = (\hat\beta, 0)$. 
Infinitesimal shifts in 
$\boldsymbol{\hat\beta}$ and $\boldsymbol{\hat\theta}$ 
can be written as 
$(d\hat\theta_i) = (d\hat\theta, \hat\theta d\phi)$
and 
$(d\hat\beta_i) = (d\hat\beta, \hat\beta d\phi)$, 
where $\phi$ denotes the azimuthal angle in the $x$-$y$ plane. 
The axial symmetry around the lens object 
implies that $\hat\theta$ and $\hat\beta$ 
are independent of $\phi$. 
Therefore, the off-diagonal terms vanish 
in the local coordinates. 
Hence, we can arrive at 
(\ref{lambdaP}) and (\ref{lambdaM}) \cite{Izumi}.

\begin{figure}
\includegraphics[width=8cm]{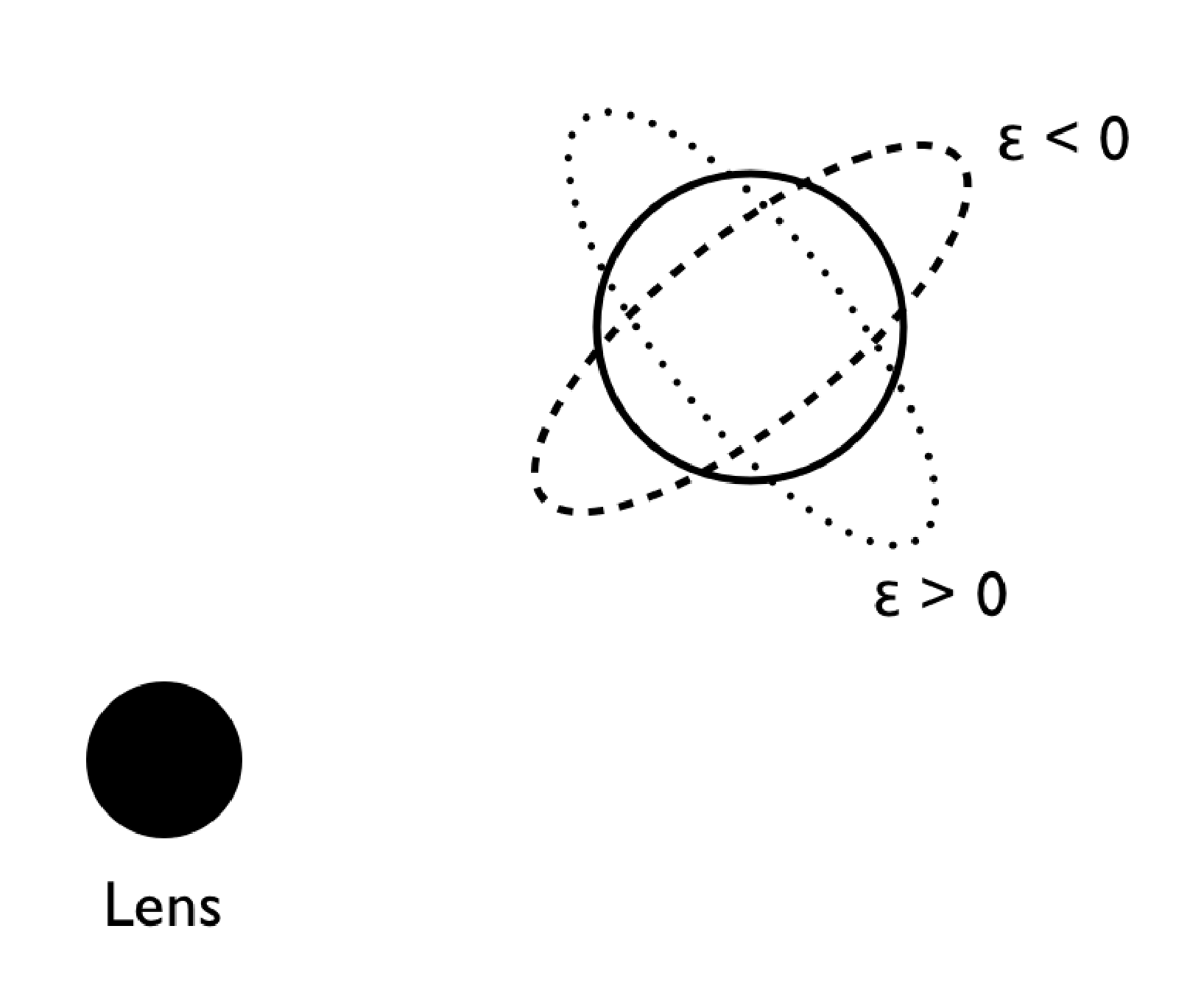}
\caption{
Schematic figure for the image distortion. 
The solid circle is an intrinsic source shape. 
The dotted ellipse is a tangential image, 
which is caused by Schwarzschild lens, as common in astronomy. 
The dashed ellipse is a radially magnified image, 
which may be caused by an exotic lens. 
}
\label{fig-shear}
\end{figure}

If and only if $n > -1$, $\lambda_{-} > \lambda_{+}$. 
Therefore, the primary image is tangentially elongated. 
Please see also Figure \ref{fig-shear}.

Eqs. (\ref{lambdaP}) and (\ref{lambdaM}) tell us 
the convergence and the shear as 
\begin{eqnarray}
\kappa &=& 1 - \frac{\lambda_{+} + \lambda_{-}}{2} 
\nonumber\\
&=&  \frac{1-n}{2}\frac{1}{\hat\theta^{n+1}} , 
\end{eqnarray}
\begin{eqnarray}
\gamma &=& \frac{\lambda_{+} - \lambda_{-}}{2} 
\nonumber\\
&=&  - \frac{1+n}{2}\frac{1}{\hat\theta^{n+1}} , 
\end{eqnarray} 
respectively. 
This expression of $\kappa$ 
agrees with Eq. (\ref{kappa}).

Next, we study the secondary image ($\hat\theta < 0$). 
By using Eq. (\ref{lenseqM}), 
one can see $\lambda_{-} > \lambda_{+}$, if and only if $n > -1$. 
Hence, also the secondary image is tangentially elongated.

Finally, we argue the dependence on the exponent $n$. 
As an example, we consider a sizably elongated case such as a giant arc 
that is often observed near the Einstein ring ($\hat\theta \sim 1$). 
For such a case, Eqs. (\ref{lambdaP}) and (\ref{lambdaM}) 
are expanded as 
\begin{eqnarray}
\lambda_{+} &=& 
(n+1) (\hat\theta - 1) 
- \frac{(n+1)(n+2)}{2} (\hat\theta - 1)^2 
+ O\left((\hat\theta - 1)^3\right) , 
\\
\lambda_{-} &=& n+1 -n(n+1) (\hat\theta - 1) 
+ O\left((\hat\theta - 1)^2\right) . 
\end{eqnarray}
Here, we used the identity $\hat\theta = 1+(\hat\theta - 1)$. 
The ratio of the tangential magnification factor to the radial one is 
\begin{equation}
\frac{\lambda_{-}}{\lambda_{+}} 
= \frac{1}{\hat\theta - 1} 
+ \left( 1 - \frac{n}{2} \right)
+ O(\hat\theta - 1) . 
\end{equation}
This ratio is corresponding to the arc shape. 
This expression suggests that, 
for the fixed observed lens position $\hat\theta$,  
the tangentially elongated shape of the lensed image 
becomes weaker, as $n$ is increased. 
This dependence on $n$ is true of also the secondary image. 

\subsection{$\varepsilon < 0$ case}
Next, we consider $\varepsilon < 0$ case. 
Eq. (\ref{lenseq}) takes 
the vectorial form as 
\begin{eqnarray}
\boldsymbol{\hat\beta}
&=& \boldsymbol{\hat\theta} 
+ \frac{\boldsymbol{\hat\theta}}{\hat\theta^{n+1}}  
\quad (\hat\theta > 0) , 
\label{lenseqP2}\\
\boldsymbol{\hat\beta}
&=& \boldsymbol{\hat\theta} 
+ \frac{\boldsymbol{\hat\theta}}{(-\hat\theta)^{n+1}}  
\quad (\hat\theta < 0) , 
\label{lenseqM2}
\end{eqnarray} 
where we use the normalization in terms of the Einstein ring radius. 

Without loss of generality, we choose $\hat\beta >0$. 
Then, Eq. (\ref{lenseqM2}) has no root for $\hat\theta < 0$, 
whereas there are at most two positive roots 
for Eq. (\ref{lenseqP2}). 

We consider a case that an impact parameter is large. 
Two images appear on the same side 
with respect to the lens position, 
though no images are caused for a small impact parameter. 
The only one image appears only when the impact parameter takes 
a critical value. 
Let us focus on the two image cases, for which the single image case 
can be studied in the limit that the impact parameter 
approaches the critical value. 

By using Eq. (\ref{lenseqP2}), 
we find 
\begin{eqnarray}
\lambda_{+} = \frac{\hat\beta}{\hat\theta} 
= 1 + \frac{1}{\hat\theta^{n+1}} , 
\label{lambdaP2}
\end{eqnarray}
\begin{eqnarray}
\lambda_{-} = \frac{d\hat\beta}{d\hat\theta} 
= 1 - \frac{n}{\hat\theta^{n+1}} . 
\label{lambdaM2}
\end{eqnarray}
If and only if $n > -1$, $\lambda_{-} < \lambda_{+}$. 
Both images are thus almost everywhere radially elongated.  
It follows that $n=1$ case, which is corresponding to the  
Schwarzschild lens with negative mass,  leads to $\kappa = 0$.

By using Eqs. (\ref{lambdaP2}) and (\ref{lambdaM2}), 
the shear can be obtained as 
\begin{eqnarray}
\gamma &=& \frac{\lambda_{+} - \lambda_{-}}{2} 
\nonumber\\
&=&  \frac{1+n}{2}\frac{1}{\hat\theta^{n+1}} . 
\end{eqnarray}

Note that a repulsive case may be related with the lensing 
by a void-like mass distribution, though 
the above calculations assume the Minkowskian background spacetime. 
There are very few galaxies in voids compared with 
in a cluster of galaxies. 
Hence, it is difficult to investigate gravity inside a void 
by using galaxies as a tracer. 
Gravitational lensing shear measurements would be another tool 
for studying voids. 
If one wishes to consider cosmological situations, 
the gravitational potential and the mass density 
might correspond to the scalar perturbation and the density contrast 
in the cosmological perturbation approach 
based on the Friedmann-Lemaitre background spacetime \cite{SEF}. 
If such a cosmological extension of the present model with $\kappa < 0$ 
is done, it can describe an underdense region called a cosmic void, 
in which the local mass density is smaller than the cosmic mean density 
and the density contrast is thus negative. 
The gravitational force on the light rays by the surrounding region 
can be interpreted as repulsive ($\varepsilon < 0)$. 
This is because the bending angle of light by the spherical void 
must be negative. 
Therefore, cosmic voids might correspond to 
a $\kappa < 0$ and $\varepsilon < 0$ case. 
Note that the positive convergence due to the cosmic mean density 
should be taken into account by the definition of the cosmological
distances but not by the gravitational lens.

Before closing this section, 
we mention whether the radial elongation can be distinguished from 
the tangential one by observations without assuming a priori 
the lens position.  
Usually, lens objects cannot be directly seen 
except for visible lens objects such as galaxies. 
In particular, exotic lens models that are discussed in this paper 
are likely to be invisible, because they are exotic. 
In the above calculations, the origin of the two-dimensional 
coordinates is chosen as the center of the lens object, 
such that the radial and tangential directions can be well defined. 
For a pair of radially elongated images ($\varepsilon < 0$), 
they should be in good alignment with each other. 
For a pair of tangentially elongated images ($\varepsilon > 0$), 
they should be parallel with each other. 
Therefore, 
the radial elongation can be distinguished in principle from 
the tangential one, 
if such a radial alignment is detected in a future observation.

\section{Astrometric lensing}
In this section, we follow Ref. \citen{Kitamura2014} 
to discuss astrometric gravitational lensing by exotic objects. 
We focus on the microlensed image centroid motions. 
Such a centroid motion is observable in astrometry. 
The astrometric microlensing by ordinary objects 
was discussed by many authors 
\cite{Walker, MY, HOF, SDR, JHP, Lewis, asada02, HL}. 

In any case of $\varepsilon > 0$ and $\varepsilon < 0$, 
the primary image position and the secondary one are 
denoted by $\boldsymbol{\hat{\theta}}_1$ 
and $\boldsymbol{\hat{\theta}}_2$, respectively, and 
the corresponding amplification factors 
are denoted by $A_1$ and $A_2$, respectively. 
Without loss of generality, 
we can choose $\hat\theta$ as $\hat\theta_1 > \hat\theta_2$. 
The center of the mass distribution plays a crucial role 
in gravitational physics.
In analogy with the mass center, 
the centroid position of the light distribution of 
a gravitationally microlensed source can be defined as 
\begin{eqnarray}
\boldsymbol{\hat{\theta}}_{\it{pc}} 
&\equiv& 
\frac{A_{1} \boldsymbol{\hat{\theta}}_{1} 
+ A_{2} \boldsymbol{\hat{\theta}}_{2}}{A_{\it{tot}}} , 
\label{pc}
\end{eqnarray}
where $A_{\it{tot}}$ means the total amplification as 
$A_{1}+A_{2}$. 
This centroid position is observable in astrometry. 
The corresponding scalar can be defined by  
$\hat{\theta}_{\it{pc}} \equiv (A_{1} \hat{\theta}_{1} 
+ A_{2} \hat{\theta}_{2}) A_{\it{tot}}^{-1}$. 
Note that $\hat{\theta}_{\it{pc}}$ is positive, 
if the centroid is located on the side same as the source 
with respect to the lens center. 

The relative displacement of the image centroid 
with respect to the source position becomes 
\begin{equation}
\delta\boldsymbol{\hat{\theta}}_{\it{pc}} 
= 
\boldsymbol{\hat{\theta}}_{\it{pc}} 
- \boldsymbol{\hat{\beta}} ,
\label{deltapc}
\end{equation}
which is often called the centroid shift. 
If intrinsic (unlensed) positions of the source 
in asymptotic regions, apparently very far from the lens, 
are known, the centroid shift also is observable. 
The scalar of the centroid shift can be defined by 
$\delta\hat{\theta}_{\it{pc}} 
\equiv \hat{\theta}_{\it{pc}} - \hat{\beta}$. 
We can see that $\delta\hat{\theta}_{\it{pc}}$ is positive, 
if $\hat{\theta}_{\it{pc}}$ is larger than $\hat{\beta}$. 

By noting the relation between the lens position and 
source trajectory in the sky, 
$\hat{\beta}$ is expressed as a function of time 
\begin{equation}
\hat{\beta}(t) = \sqrt{\hat{\beta}_0^2 + {(t -t_0)^2/t_{\it{E}}}^2} , 
\label{betat}
\end{equation}
where $\hat{\beta}_0$ denotes the impact parameter of 
the source trajectory and $t_0$ denotes the passage time at closest approach. 
Here, we assume that the source is in uniform linear motion. 
This assumption is reasonable for a case that 
the observational time scale is much longer than 
the time scale of the galactic rotation motion. 
For its simplicity, we choose $t_0$ as $t_0 = 0$. 
The Einstein radius crossing time is denoted as $t_{\it{E}}$, 
which is defined by
\begin{equation}
t_{\it{E}} = R_{\it{E}} / v_{\it{T}},  
\label{eqn:te}
\end{equation}
where $v_{\it{T}}$ denotes the transverse velocity of the lens 
relative to the source.  
Please see Ref. \citen{Kitamura2014} for detailed discussions and numerical figures 
by exotic lenses.

\section{Gravitational time delay}
In this section, we follow Ref. \citen{Nakajima2014} 
to study the gravitational time delay 
of light by using the inverse-power exotic lens model. 
\subsection{Time delay of a light signal} 
Again, we use Eq. (\ref{ds}) as the spacetime metric. 
By noting $ds^2 = 0$ for a light signal, 
we obtain 
\begin{equation}
\left(\frac{dr}{dt}\right)^2
=\left(1-\dfrac{\varepsilon_1}{r^n}\right)
\left(1-\dfrac{\varepsilon_2}{r^n}\right)
\left[1-\dfrac{b^2}{r^2}\left(1-\dfrac{\varepsilon_1}{r^n}\right)\right] 
+O(\varepsilon_1^2, \varepsilon_2^2, \varepsilon_1 \varepsilon_2) .
\label{orbit}
\end{equation}
The impact parameter $b$ is related with 
the closest approach $r_0$ as 
\begin{equation}
b^2=\cfrac{r_0^2}{1-\cfrac{\varepsilon_1}{r_0^n}}.
\label{b}
\end{equation}
By using Eq. (\ref{orbit}), 
the time-of-flight of a photon from the source (denoted by $S$) 
to the receiver (denoted by $R$) is computed as 
\begin{eqnarray}
t(S \to R)
&\equiv& \int_{r_S}^{r_R} dt 
\nonumber\\
&=&\int_{r_S}^{r_R} 
\left(1-\dfrac{r_0^2}{r^2}\right)^{-\frac{1}{2}}
\left(
1-\dfrac{\dfrac{r_0^2}{r^2}\dfrac{\varepsilon_1}{r_0^n}
\left(1-\dfrac{r_0^n}{r^n}
\right)}
{1-\dfrac{r_0^2}{r^2}}
-\dfrac{\tilde{\varepsilon}}{r^n}
\right)^{-\frac{1}{2}}dr
\nonumber\\
&&
+O(\varepsilon_1^2, \varepsilon_2^2, \varepsilon_1 \varepsilon_2) ,
\label{traveltime}
\end{eqnarray}
where we $\tilde{\varepsilon}\equiv\varepsilon_1+\varepsilon_2$.

We compare Eq. (\ref{traveltime}) with the time-of-flight 
in the Minkowskian spacetime to 
find the time delay at the linear order as 
\begin{equation}
\delta t = 
\frac{1}{r_0^{n-1}}
\int_{\Psi_{S}}^{\Psi_{R}} 
\left(\dfrac{\varepsilon_1(1-\cos^n\Psi)}{\sin^2\Psi}
+\tilde{\varepsilon}\cos^{n-2}\Psi\right)d\Psi , 
\label{delay}
\end{equation}
where $\Psi_R$ and $\Psi_S$ are corresponding to 
the direction from the lens to the receiver and 
that to the source, respectively.

The integral in Eq. (\ref{delay}) for general $n$ 
is not necessarily expressed in terms of elementary functions. 
Numerical calculations of Eq. (\ref{delay}) are thus needed. 
However, in particular, if $n$ is an integer, 
the integration of powers of trigonometric functions 
can be simplified \cite{GR}. 
For $n=1$ for instance, Eq. (\ref{delay}) is simplified as  
\begin{equation}
\delta t_1
=\varepsilon_1\left[
\dfrac{r_R}{x_R}+\dfrac{r_S}{x_S}
-\left(\dfrac{r_0}{x_R}+\dfrac{r_0}{x_S}\right)
\right]
+\tilde{\varepsilon}\left[
2\ln\dfrac{(r_R+x_R)(r_S+x_S)}{r_0^2}\right] ,
\label{delay-n1}
\end{equation}
where we define $x_R \equiv \sqrt{r_R^2 - r_0^2}$ and 
$x_S \equiv \sqrt{r_S^2 - r_0^2}$. 
It follows that Eq. (\ref{delay-n1}) recovers the Shapiro delay formula 
if we assume $\varepsilon_1 = \varepsilon_2$ and 
the parameters are the same as the Schwarzschild radius.

For $n=2$, Eq. (\ref{delay}) becomes 
\begin{equation}
\delta t_2
=\dfrac{\varepsilon_1+\tilde{\varepsilon}}{r_0}
\left(\arccos\dfrac{r_0}{r_R}+\arccos\dfrac{r_0}{r_S}\right).
\label{delay-n2}
\end{equation}

Next, we consider the case of an even integer ($n=2p$), 
where $p$ means a positive integer. 
We thus obtain 
\begin{eqnarray}
\delta t_{2p}
&=&
\dfrac{\varepsilon_1}{c r_0^{2p-1}}
\nonumber\\
&&
\times
\left\{
    - \cot \Psi_R
    + \dfrac{\cos^{2p+1} \Psi_R}{\sin \Psi_R}
    + \dfrac{(2p-1)!!}{(2p-2)!!} \sin \Psi_R
    \sum^{p-1}_{r=0} \dfrac{(2p-2r-2)!!}{(2p-2r-1)!!} \cos^{2p-2r-1} \Psi_R 
\right.
\nonumber
\\
&\ &
~~~~~~~~~~~~~
\left.
+ \dfrac{(2p-1)!!}{(2p-2)!!} \Psi_R 
\right.
\nonumber
\\
&\ &
~~~~~~~~~~~~~
\left.
    + \cot \Psi_S
    - \dfrac{\cos^{2p+1} \Psi_S}{\sin \Psi_S}
    - \dfrac{(2p-1)!!}{(2p-2)!!} \sin \Psi_S
    \sum^{p-1}_{r=0} \dfrac{(2p-2r-2)!!}{(2p-2r-1)!!} \cos^{2p-2r-1} \Psi_S
\right.
\nonumber
\\
&\ &
~~~~~~~~~~~~~
\left.
- \dfrac{(2p-1)!!}{(2p-2)!!} \Psi_S 
\right\}
\nonumber
\\
&\ &
+
\dfrac{\tilde{\varepsilon}}{c r_0^{2p-1}}
\nonumber\\
&&
\times
\left\{
      \dfrac{(2p-3)!!}{(2p-2)!!} \sin \Psi_R
      \sum^{p-2}_{r=0} \dfrac{(2p-2r-4)!!}{(2p-2r-3)!!} \cos^{2p-2r-3} \Psi_R
      + \dfrac{(2p-3)!!}{(2p-2)!!} \Psi_R
\right.
\nonumber
\\
&\ &
~~~~~~~~~~~~~
\left.
      - \dfrac{(2p-3)!!}{(2p-2)!!} \sin \Psi_S
      \sum^{p-2}_{r=0} \dfrac{(2p-2r-4)!!}{(2p-2r-3)!!} \cos^{2p-2r-3} \Psi_S
      - \dfrac{(2p-3)!!}{(2p-2)!!} \Psi_S
\right\} , 
\label{delay-neven}
\end{eqnarray}
where the subscript $2p$ means the case of $n=2p$ and 
$(2p-1)!!$ denotes $(2p-1)(2p-3) \cdots 1$. 

Next, we examine the case of $n=2p+1$. 
Eq. (\ref{delay}) becomes 
\begin{eqnarray}
\delta t_{2p+1}
&=&
\dfrac{\varepsilon_1}{c r_0^{2p}}
\nonumber\\
&&
\times
\left\{
    - \cot \Psi_R
    + \dfrac{\cos^{2p+2} \Psi_R}{\sin \Psi_R}
    + \dfrac{(2p)!!}{(2p-1)!!} \sin \Psi_R
      \sum^p_{r=0} \dfrac{(2p-2r-1)!!}{(2p-2r)!!} \cos^{2p-2r} \Psi_R
\right.
\nonumber
\\
&\ &
~~~~~~~~~~
\left.
    + \cot \Psi_S
    - \dfrac{\cos^{2p+2} \Psi_S}{\sin \Psi_S}
    - \dfrac{(2p)!!}{(2p-1)!!} \sin \Psi_S
      \sum^p_{r=0} \dfrac{(2p-2r-1)!!}{(2p-2r)!!} \cos^{2p-2r} \Psi_S
\right\}
\nonumber
\\
&\ &
+
\dfrac{\tilde{\varepsilon}}{c r_0^{2p}}
\nonumber\\
&&
\times
\left\{
      \dfrac{(2p-2)!!}{(2p-1)!!} \sin \Psi_R
      \sum^{p-1}_{r=0} \dfrac{(2p-2r-3)!!}{(2p-2r-2)!!} \cos^{2p-2r-2} \Psi_R
\right.
\nonumber
\\
&\ &
~~~~~~~~~~
-
\left.
      \dfrac{(2p-2)!!}{(2p-1)!!} \sin \Psi_S
      \sum^{p-1}_{r=0} \dfrac{(2p-2r-3)!!}{(2p-2r-2)!!} \cos^{2p-2r-2} \Psi_S
\right\} . 
\label{delay-nodd}
\end{eqnarray}

Up to this point, integrations have been done 
without assuming any approximation. 
In astronomical observations, 
the closest distance of light is usually much smaller than 
the distance from the lens to the source 
and that from the lens to the observer, 
such that we can use the limit as $r_S/r_0 \to \infty$ 
and $r_R/r_0 \to \infty$, 
namely $\Psi_R \to \pi/2$ and $\Psi_S \to - \pi/2$, 
to simplify some expressions. 
 
For $n=2p$, Eq. (\ref{delay-neven}) approaches 
\begin{eqnarray}
    \delta t_{2p} \to \frac{\pi}{c} \dfrac{(2p-3)!!}{(2p-2)!!} 
\dfrac{2p \varepsilon_1 + \varepsilon_2}{r_0^{2p-1}} . 
\label{delay-neven-limit}
\end{eqnarray}
For $n=2p+1$ case, Eq. (\ref{delay-nodd}) becomes approximately 
\begin{eqnarray}
    \delta t_{2p+1} \to \frac{2}{c} \dfrac{(2p-2)!!}{(2p-1)!!} 
\dfrac{(2p + 1) \varepsilon_1 + \varepsilon_2}{r_0^{2p}} , 
\label{delay-nodd-limit}
\end{eqnarray} 
where we use $[\cos\Psi]^0 = 1$. 
We should note that most of terms in Eqs. (\ref{delay-neven}) 
and (\ref{delay-nodd}) vanish in 
Eqs. (\ref{delay-neven-limit}) and (\ref{delay-nodd-limit}) 
because $\cos(\pm \pi/2) = 0$ in the above limit. 
According to Eqs. (\ref{delay-neven-limit}) and
(\ref{delay-nodd-limit}),  
the time delay $\delta t$ is proportional to 
$n \varepsilon_1 + \varepsilon_2 = \varepsilon$ 
and also to $r_0^{-(n-1)}$, 
if the receiver and source are very far from the lens. 
The sign of the time delay 
is the same as that of the deflection angle of light. 
The time delay $\delta t$ for $\varepsilon > 0$ 
is a downward-convex function of $r_0$ almost everywhere 
except for $r_0 = 0$, 
whereas $\delta t$ for $\varepsilon < 0$ is convex upward.

\subsection{Frequency shift}
We discuss the frequency shift that is caused by the time delay. 
In astronomy, 
the exact time of the emission of light is usually unknown. 
Therefore, the arrival time delay is not a direct observable  
for a single image, 
whereas the arrival time difference between multiple images 
(e.g. distant quasars) are observable in astronomy \cite{SEF}. 
Note that the round-trip time of a light signal 
is directly observable only for a spacecraft experiment 
such as Voyager and Cassini. 
On the other hand, the frequency shift induced by the time delay 
becomes a direct observable even though the emission time is unknown, 
if the emitter of a light signal 
moves with respect to the lens object \cite{Cassini,Asada2008}. 

We can define the frequency shift ratio $y$ due to the time delay 
as \cite{Cassini,Asada2008}
\begin{eqnarray}
y &\equiv& \frac{\nu(t) - \nu_0}{\nu_0} 
\nonumber\\
& =& - \dfrac{d(\delta t)}{dt} , 
\label{y}
\end{eqnarray} 
where $\nu_0$ denotes the intrinsic frequency of light 
and $\nu(t)$ means the observed one at time $t$. 

If the emitter of light approaches the lens, $y < 0$. 
If it moves away, $y > 0$. 
A general expression of $y$ may be pretty long, 
though it is simplified for the case of an integer $n$. 
First, we take a derivative of 
Eqs. (\ref{delay-neven}) 
and (\ref{delay-nodd}) with respect to time, 
where we use 
\begin {eqnarray}
    \dfrac{d \Psi_I}{dt} = 
- \dfrac{1}{\sqrt{r_I^2 - r_0^2}} \dfrac{dr_0}{dt} ,   
\end{eqnarray}
for $I = R, S$. 
Next, we take the far limit as 
$\Psi_R \to \pi/2$ and $\Psi_S \to - \pi/2$. 

For an even integer $n=2p$, 
the frequency shift is computed as 
\begin{eqnarray}
    y_{2p} = \frac{\pi}{c} \dfrac{(2p-1)!!}{(2p-2)!!} 
\dfrac{\varepsilon}{r_0^{2p+1}} v^2 t . 
\label{y2p}
\end{eqnarray}
For an odd integer $n=2p+1$, 
it becomes 
\begin{eqnarray}
    y_{2p+1} = \frac{2}{c} \dfrac{(2p)!!}{(2p-1)!!} 
\dfrac{\varepsilon}{r_0^{2p+2}} v^2 t .  
\label{y2p+1}
\end{eqnarray} 

Eqs. (\ref{y2p}) and (\ref{y2p+1}) become  
$y \propto t/r_0^{n+1} \propto t^{-n}$ for large $|t|$. 
Therefore, the frequency shift for large $n$ 
decreases more rapidly as $r_0$ increases.

\subsection{Possible parameter ranges in pulsar timing method} 
There are two important quantities in time delay measurements. 
One quantity is the maximum amplitude of the time delay,
and the other is the time duration. 
{}From Eqs. (\ref{delay-neven-limit}) and (\ref{delay-nodd-limit}), 
the size of the time delay is roughly 
\begin{equation}
\delta t \sim \frac{1}{c} \frac{\varepsilon}{r_0^n} , 
\end{equation}
and the duration of the time delay is approximately 
\begin{equation}
T_{td} \sim \cfrac{\delta t}{\left( \cfrac{d(\delta t)}{dt}\right)} 
\sim \frac{r_0}{v} .
\end{equation}

Let us assume that $v$ is of the order of the typical rotational velocity
in our galaxy, 
because currently we have no reliable theories 
on the motion of the exotic lens. 
We consider the pulsar timing measurements 
which are currently done a few times per year for each pulsar.
The root-mean-squared residuals of the timing arrival dispersions are 
around 100 - 1 nanoseconds (ns) that are depending on pulsars 
\cite{Pulsar,Yunes}.
The closest approach is typically 
\begin{equation}
r_0 \sim 40 \mbox{AU}  
\left(\frac{T_{td}}{1 \mbox{year}}\right)
\left(\frac{v}{200 \mbox{km/s}}\right) .
\end{equation}
Roughly speaking, the exotic lens potential is 
\begin{equation}
\frac{\varepsilon}{r_0^n} \sim 10^{-11}
 \left(\frac{\delta t}{100 \mbox{ns}}\right) .
\end{equation}
This case $\delta t \sim 100$ ns 
is corresponding to the frequency shift of $y \sim 10^{-15}$. 
The accuracy for $y \sim 10^{-15}$ 
has been already achieved for a milli-second pulsar 
measurement \cite{Pulsar,Yunes}. 

Next, we discuss a possible constraint on the number
density of the exotic lens models.
One event would be detectable, 
if a lens object entered a volume of a cylinder as 
$V \sim \pi r_0^2 D_S$, where $D_S$ denotes the distance of 
the source, presumably pulsar, from the observer.
The effective survey volume is proportional to both the number of the
observed pulsars (denoted as $N_p$) 
and the total observation time of a pulsar timing project 
(denoted as $T_{pt}$). 
Hence, if no exotic time delay is detected, 
the number density of the lens objects $\Omega_L$ 
can be constrained as
\begin{equation}
\Omega_L < 10^3 \mbox{pc}^{-3} 
\left( \frac{40 \mbox{AU}}{r_0} \right)^2 
\left( \frac{1 \mbox{kpc}}{D_S} \right)
\left( \frac{10 \mbox{year}}{T_{pt}} \right) . 
\end{equation}
This constraint is very weak. 
However, this constraint may be interesting, 
because models for $n>1$ are massless 
at the spatial infinity and thus it is extremely difficult to probe 
these exotic objects by other astronomical observations 
such as stellar motions, galactic rotation. 
Please see e.g. Ref. \citen{DA} for the gravitational time delay 
by naked singularities.

\section{Conclusion}
First, we discussed the light propagation 
in the Ellis wormhole lens in detail. 
We considered several attempts to put a constraint on 
the Ellis wormhole by using the existing astronomical observations. 
Next, we examined a unified model for describing simply 
exotic lenses, for which the spacetime metric 
is expressed as a function in the inverse powers of the distance. 
As observational implications, 
the magnification, shear, photo-centroid motion 
and time delay in this lens model were studied. 

There are two future directions along the present issue. 
One direction is to seek a more general form of an exotic gravitational lens. 
By using the spacetime model, observational implications should be 
discussed. 
The other direction is to 
make use of a much bigger observational data sets 
for probing such an exotic gravitational lens.

\section*{Acknowledgments}
We would like to thank Fumio Abe 
for the very fruitful discussions, 
especially his seminar talk at Hirosaki University, 
by which our work is strongly motivated. 
We are grateful to Naoki Tsukamoto 
for the useful comments. 
We thank Ryuichi Takahashi for the useful discussion, especially 
on the negative mass lens. 
We thank Takao Kitamura, Koki Nakajima, 
Marcus Werner, Koji Izumi, Chisaki Hagiwara 
for stimulating discussions. 
We are grateful to Yuuiti Sendouda, Tomohiro Harada, Ken-ichi Nakao, 
Naoteru Gouda, Masumi Kasai for the fruitful conversations.  
This work was supported 
in part by Japan Society for the Promotion of Science 
Grant-in-Aid for Scientific Research, 
No. 26400262, No. 17K05431 and 
in part by by Ministry of Education, Culture, Sports, Science, and Technology,  
No. 15H00772 and No. 17H06359.



\begin{thebibliography}{0}
\bibitem{Nandi-2}
We should note that 
Eq. (54) in \citen{Nandi} follows an uncommon notation. 
Their argument of the first kind Elliptic integral 
is $k^2$. 
This is a square of the argument for $K(k)$ in the common notation.
\end{thebibliography}
\end{document}